\documentclass[10pt]{article}
\usepackage{amsmath}
\usepackage{epsfig}

\renewcommand{\a}{\alpha} 
\renewcommand{\b}{\beta}

\newcommand{\g}{\gamma}

\newcommand{\s}{\sigma}
\renewcommand{\S}{\Sigma} 

\renewcommand{\t}{\tau}

\newcommand{\tS}{\tilde{S}}

\newcommand{\pd}{\partial}

\newcommand{\half}{\frac{1}{2}}

\begin{document}

\twocolumn
\author{Jiri Hoogland\footnote{jiri@cwi.nl} 
  and Dimitri Neumann\footnote{neumann@cwi.nl}\\
  CWI,\\ P.O.~Box 94079, 1090 GB  Amsterdam, The Netherlands\\[5mm]
  Daniel Bloch\footnote{Daniel.Bloch@dresdnerkw.com}\\
  Dresdner Kleinswort Wasserstein,\\
  20 Fenchurch Street, London, EC3P 3DB United Kingdom}

\title{\textbf{Converting the reset}}
\maketitle

\thispagestyle{empty}
\begin{abstract}
  We give a simple algorithm to incorporate the effects of resets in
  convertible bond prices, without having to add an extra factor to
  take into account the value of the reset. Furthermore we show that
  the effect of a notice period, and additional make-whole features,
  can be treated in a straightforward and simple manner.  Although we
  present these results with the stockprice driven by geometric
  Brownian and a deterministic interest term structure, our results
  can be extended to more general cases, e.g. stochastic interest
  rates.
\end{abstract}

\section{Introduction}
\label{sec:introduction}

In recent years many Japanese companies have issued convertible bonds
with reset-features. At specific reset-dates the conversion ratio
\footnote{The conversion ratio is the amount of stock the bond is
  converted into. The conversion value is the conversion ratio times
  the spotprice of the stock. The conversion price is the face value
  of the bond divided by the conversion ratio.}
of the convertible bond (CB) is reset such that the conversion value
is equal to the face value in cash. Typically there is one reset and
the reset of the conversion price is only allowed to be downward.

The reset feature makes the contract more attractive, since it
protects the holder of the bond against drops in the value of the
underlying stock, but it is also more difficult to price. The usual
approach in the literature~\cite{Wilmott98,Nelken98} is to introduce
an extra factor to take into account the reset-feature. The CB is then
priced by solving a two- or three-dimensional PDE or using a
multi-dimensional tree algorithm. This is time-consuming and
error-prone.

In this article, we show that, in general, resets can be simply taken
into account as jump conditions at the reset dates. This prevents the
introduction of the extra factor, which greatly simplifies and
improves the speed of the algorithm.

Furthermore we show that the effects of a notice period can be simply
incorporated in the pricing problem by using an effective call
boundary value, that follows from the remaining optionality for the
holder to convert during the holding period.

Finally we show that, in a deterministic interest rate setting, the
resettable convertible, without callable and puttable features, can be
related to an American discrete lookback option.

The outline of this article is as follows. In
Section~\ref{sec:definition-problem} we define the convertible bond
(CB) in terms its boundary conditions etc. In Section~\ref{sec:model}
we describe the model we use to price the CB.
Section~\ref{sec:notice-period} and
section~\ref{sec:resettable-features} shows how the notice period and
resets can be incorporated in a numerical pricing scheme. In
Section~\ref{sec:nice-relation} we show that a change of numeraire
provides an alternative way of understanding reset features in terms
of lookbacks. Section~\ref{sec:results} discusses implementation of
the scheme and compares our results to other results in the literature
and market prices. In Section~\ref{sec:conclusions} we conclude and
discuss possible extensions of our results to more complex settings.

\section{Definition of the problem}
\label{sec:definition-problem}

In this article we use the formulation of option pricing in terms of
self-financing objects, which we call tradables. In particular we will
not use cash as this is not self-financing, instead we use zero-coupon
bonds. More details can be found in ~\cite{HooglandNeumann99a}.

A convertible bond (CB) is a coupon-paying corporate bond $B$ of
maturity $T$, which has the embedded optionality to convert into a
specified number, the conversion ratio $k$, of underlying stock $S$.
The conversion is optimal, when the value $V$ of the CB is less than
the conversion value $kS$
\[
V(S,B,t)\le k S
\]
The conversion ratio may depend on time and the paths of the
underlyings. When the CB is in-the-money, it behaves much like the
underlying stock
\[
V(S,B,t) \approx k S
\]
where $S$ has $N_d$ dividend payments $d_iS$ at dates $t_{d,i}$.  When
the CB is out-of-the-money it behaves like a coupon-paying corporate
bond
\[
V(S,B,t) \approx B = FP_T+\sum_{i:t\le t_{c,i}\le T} c FP_{t_{c,i}}
\]
where $P_s$ is a discount bond worth $1\$ $ at maturity $s$, the face
value is denoted by $F$, and there are $N_c$ coupon payments $c$
(denoted as a fraction of the face value) at dates $t_{c,i}$.  If the
CB has not been converted, the payoff at maturity will be given by
\[
V(S,B,T)=B
\]
When there is mandatory conversion, this changes to $V(S,B,T)=kS$.  A
typical CB also has callable and puttable features.  The callable
feature allows the bond to be called by the issuer, when the CB price
rises above a level $M_C$ of cash. This limits the potential loss of
the issuer. The puttable feature allows the holder to redeem an amount
of cash when the CB price drops below $M_P$ of cash. These features
translate into the following bounds on the price of the convertible
\[
M_P P_t\le V(S,B,t)\le M_C P_t
\]
Here we write $P_t=P_t(t)$ to denote that the constraint is in terms
of cash at time $t$, $P_t=1$ in units of the currency.  The puttable
and callable features may be time-depen\-dent.  Typically a contract
is continuously callable, while the puttable feature is active only at
a discrete set of times.

Also when a contract is called, the holder typically has the freedom
to convert during a specified {\it notice} period after calling of the
contract.  This notice period is typically of the order of a few
months. We will come back to the effects of the notice period in
Section~\ref{sec:notice-period}.

Furthermore the conversion ratio may be reset at $N_r$ prespecified
dates $t_{r,i}$ according to specific rules. The reset feature is in
general considered a difficult problem to deal with since it
introduces a path-depen\-dancy in the pricing of the contract. It
turns out however that the path-dependancy is of the soft sort,
similar to the case of barrier- and lookback- contracts. This means
that we can move the path-dependancy to the jump conditions, keeping
the pricing relatively simple.

\section{Modelling convertible bonds}
\label{sec:model}

The modelling of a CB is a relatively complex issue, due to its
sensitivity to interest rates, credit risk, and stock volatility.  A
proper model should, of course, include all these features. The more
realistic case with a stochastic interest-rate model and credit risk
will be discussed in another article.  In the present article we will
keep things simple, since we want to focus on how to deal with resets
in the pricing problem. Also we will consider a complete market with
the usual assumptions; we are allowed to short stocks, have no
transaction costs etc. When there is a deterministic relation between
discount bonds with different maturities we have assuming, for
simplicity, continuous compounding and constant rate $r$,
\[
P_s(t)= e^{r(T-s)}P_T(t)
\]
This implies that we can rewrite every occurance of a discount bond
with a maturity different from $T$ into one maturing at $T$ times some
deterministic time-dependent factor. This simplifies the discussion
considerably.
Cashflows at any time $s$ can be converted easily into discount bonds
maturing at time $T$ using the above relation. For example, the coupon
bond at time $t$ becomes
\[
B = FP_T +\sum_{i:t\le t_{c,i}\le T} c F e^{r(T-t_{c,i})}P_T
\]
Here and in the following we will use the shorthand notation $P\equiv
P_T$. Since the contract defines the exchange between corporate bonds
and stock, credit risk is involved with every occurance of a corporate
discount bond.  The rate $r$ should therefore be understood as being
the risky rate, including a credit spread, for that particular
corporate discount bond. This is clearly a rather simple approach to
credit risk since it does, for example, not incorporate the
correlation between a drop in the stockprice and an increase in the
credit spread. Still by formulating the pricing problem in terms of
the underlying instruments it should provide a reasonable first order
approximation to incorporating credit risk. We will come back to the
credit risk issue in more detail in the second article. The stockprice
is modelled by a geometric Brownian motion:
\[
\frac{dS}{S} = \s\, dW+\ldots
\]
w.r.t. the discount bond-price $P$ maturing at $T$ and $\s$ denotes
the deterministic volatility function\footnote{ It is of course easy
  to introduce a local volatility function: $\s(t)\to\s(S,P,t)$.  We
  will not deal with this case in the present work.} of the stock.
The $W$ is Brownian motion (under the forward-$T$ measure) and the
dots denote irrelevant drift terms.

Due to the deterministic interest rate term structure, the only two
relevant variables in the problem become $S$ and $P$.  The value of
the CB at time $t$ is therefore denoted by $V(S,P,t)$ and it satisfies
the following PDE (see Ref.~\cite{HooglandNeumann99a} for details)
\begin{equation}
  \label{eq:1}
  \bigg(\pd_t+\half\s^2S^2\pd_S^2\bigg)V(S,P,t) = 0
\end{equation}
which is the symmetric version of the Black-Scholes equation. They
both lead to the same answer, but the above equation is much more
convenient to use both in deriving analytic results as well as in
solving pricing equations numerically. The convertible price should
satisfy the following constraints
\[
V(S,P,t)\ge k S
\]
and
\[
M_P e^{r(T-t)}P\le V(S,P,t)\le M_C e^{r(T-t)}P
\]
Note that these constraints may be time-dependant, e.g. CB's are
typically call-protected during the first years. The boundary
conditions are as follows. At maturity we have the payoff
\[
V(S,P,T) = (F+cF)P
\]
When there is a mandatory conversion at maturity this changes
accordingly to
\[
V(S,P,T) = kS
\]
For $S\to\infty$ we have the condition
\[
V(S,P,t) \to k S
\]
For $S\to0$ we have the condition
\[
V(S,P,t) = FP+\sum_{i:t\le t_{c,i}\le T} c F e^{r(T-t_{c,i})}P
\]
At the coupon-payment dates $t_{c,i}$ we have the jump-conditions
\[
V(S,P,t_{c,i}^-) = V(S,P,t_{c,i})+c F e^{r(T-t_{c,i})} P
\]
These are trivial to implement.  At the dividend-payment dates
$t_{d,i}$ we have the jump-conditions
\[
V(S,P,t_{d,i}^-) = V(S(1+d_i),P,t_{d,i})
\]
The jump condition at ex-dividend dates can be removed by using a
different variable instead of $S$ as we will show now. At a dividend
payment date $t_{d,i}$ we have
\[
S(t_{d,i})=\frac{S(t_{d,i}^-)}{1+d_i}
\]
Now introduce a new variable $\tS$, which is proportional to the
self-financing portfolio, hence a tradable, consisting of the stock
together with its dividends.  In the case of known stock dividends
$d_iS$, this new tradable $\tS$ is just proportional to $S$ itself,
where the factor of proportionality, $D\le 1$, jumps at dividend
payments. The variable is normalized such that it coincides with $S$
at maturity.
\[
\tS(t) \equiv S(t)\prod\limits_{i:t\le t_{d,i}\le T}
\big(1+d_i\big)^{-1} \equiv S(t) D(t)
\]
Note that $\tS$ just follows geometric Brownian motion without jumps.
In terms of $\tS$, we do not need to include jump conditions for the
dividends in the PDE at all. The effect of dividends\footnote{The case
  of continuous dividends can be treated in a similar manner, but we
  do not have to care about any jump-conditions. In the continuous
  case we have $D(t)=e^{-q(T-t)}$.}  moves entirely to the boundary
conditions.  In the present case this means that we have to change the
conversion condition to
\[
V(S,P,t) \equiv\tilde{V}(\tS,P,t) \le \frac{k\tS}{D(t)}
\]
Here the tilde is used to indicate the price as a function of the
tradable $\tS$.  Since $D(t)$ is an increasing function of time, it
increases the incentive to early convert the CB. In the case of
discrete dividends, the optimal conversion will be just before an
ex-dividend date, since the conversion value decreases at that moment.
The usefullness of this parametrization extends clearly beyond the
present case. Similar considerations allows us to treat cash-dividends
in a consistent manner~\cite{HooglandNeumann00a}.

\section{The notice period effect}
\label{sec:notice-period}

The notice period $\t_n$ is normally not taken into account in the
numerical evaluation. At first sight it looks like a complex boundary
condition.  But a closer look reveals that it can be dealt with in a
straightforward way. The CB may be called when
\begin{equation}
  \label{eq:2}
  V(S,P,t) \ge M_C e^{r(T-t)}P
\end{equation}
When the value of the CB is above this boundary, the issuer is allowed
to call the CB. In the case of calling, the holder is still allowed to
convert during the notice period. So the holder has in fact another CB
with value $V'$, a lifetime equal to $\t_n$, conversion ratio $k$, and
face value $M_C\exp(r(T-t_n))P$, where $t_n$ is the time of calling.
But there are no more callable or puttable features to take care of.
This fact allows one to compute the value $V'$ (semi-)analytically.

The conversion, after calling, depends completely on the holder of the
contract. Since the typical period is between one and four months,
there will be at most one coupon and one dividend payment during the
notice period. If there are no coupon- and dividend payments during
the notice period, there is no reason to convert early by the holder.
The CB is just valued as an European call plus bond. In the case of
one dividend payment, the only optimal early conversion is just before
the ex-dividend date. In that case the CB is valued by a bivariate
integral. Again this can also be incorporated without too much work.
Finally coupon payments are trivial to deal with. So in all cases one
is able to compute $V'$. Since $V'$ has time value it will be above
the call boundary value.

The issuer is allowed to call when Eq.~(\ref{eq:2}) holds. Whether it
is optimal to call depends on the value of $V$ and $V'$. When called
the contract will be worth $V'$ to the holder. So for the issuer
calling will depend on
\[
V(S,P,t_-) = \min\big(V(S,P,t),V'(S,P,M_C,t)\big)
\]
When $V<V'$ on the RHS, there is clearly no reason to call. On the
other hand, when $V>V'$ on the LHS it is optimal to call. The effect
of the notice period translates to an effective call boundary value
$V'$; in a numerical scheme we have to replace the condition of
calling, Eq.~(\ref{eq:2}), by
\[
V(S,P,t)\le V'(S,P,M_C,t)
\]
Note that when $\t_n\to0$, the time value of $V'$ goes to zero too and
the call boundary value will coincide with the original call value.

Often CB's contain a make-whole feature, which compensates the holder
by paying an extra amount of cash or increasing the conversion ratio,
when the contract is called. Clearly such features can be simply
incorporated by adjusting $V'$ appropriately.

By taking into account the notice period one adds extra optionality
for the holder of the CB. Hence the contract will become slightly more
expensive. The numerical effect is however small for typical notice
periods and we will neglegt it in the numerical computations of
Section~\ref{sec:results}.

\section{Resetting conversion ratios}
\label{sec:resettable-features}

When the stockprice drops it becomes less attractive to convert the
CB. Here enters the reset, which allows the conversion price to be
refixed in order to get parity around par during the lifetime of the
contract. The reset increases the value of the contract for the holder
as it improves the probability of conversion even with falling stock
prices. For the issuer resets are also attractive because the higher
price allows them to lower the coupon rate.  This also makes it
understandable why Japanese corporations were especially interested in
such contracts. In the deteriorating Japanese market resets were added
to CB's to increase their attractiveness.

For example, a CB with a downward reset protects the holder from large
drops in value of the underlying stock by resetting the conversion
price $F\$/k_i$ downward such that the conversion value $k_iS$ is
at-the-money with the face value $F\$= FP_{t_{r,i}}(t_{r,i})$ at
prespecified dates $t_{r,i}$ ($i=1,\ldots,n\equiv N_r$) during the
lifetime of the contract. Since a downward reset of the conversion
price increases the conversion ratio and hence increases the dilution,
the reset is often floored and capped by multiples of previous
conversion prices.  In practice many of the contracts have only one
reset, a few have two or even three resets.

Since resets, in general, introduce path-depen\-dance in the contract,
the usual approach to price resettable CB's is to introduce an extra
degree of freedom to take care of that fact. In many cases found in
practice this turns out to be an unnecessary complication. Instead the
resets can be treated as advanced jump-conditions at the reset-dates.
This makes the pricing of resettable convertibles not more complicated
than other types of convertibles.

In Ref.~\cite{MerrillLynch99} two types of resets are considered.  The
first case is the step-down reset, which is simply a convertible with
a deterministic time-dependant conversion price.  Clearly, this does
not require any additional trickery to price above standard CB's.  So
we will not discuss this type of contract any further.

In the second case the conversion price is reset in order to get
parity around par
\footnote{In practice the reset depends on the arithmetic average of
  the closing stockprices over a specified period before and at the
  reset date. We will just take the stockprice at the reset-date.}
, but capped and floored by multiples of previous values of the
conversion price.  In this article we will focus on the reset rule,
given in Eq.~(\ref{eq:3}). In terms of the conversion ratio, it boils
down to the following:
\begin{equation}
  k_i=
  \max\big(\a k_{i-1},
  \min\big(\b k_{i-1}, \frac{ F P_{t_{r,i}}(t_{r,i})}{S(t_{r,i})}\big)\big)
  \label{eq:3}
\end{equation}
where the conversion ratio is capped and floored by multiples of the
{\it previous} conversion ratio $k_{i-1}$, with $\a\le1$ and $\b\ge1$
typically.

The reset rule, most often found in practice, is as follows:
\begin{equation}
  k_i=
  \max\big(\a k_{i-1},
  \min\big(\b k_0, \frac{ F P_{t_{r,i}}(t_{r,i})}{S(t_{r,i})}\big)\big)
  \label{eq:4}
\end{equation}
Here the conversion ratio is capped by a multiple of the previous
conversion ratio $k_{i-1}$ from above and floored by a multiple of the
initial conversion ratio $k_0$ from below. Contracts with the reset
rule Eq.~(\ref{eq:4}) can be valued using our method, when there is
{\it only} one reset. In that case, Eqs.~(\ref{eq:3}) and (\ref{eq:4})
are identical.  With deterministic interest rates, we then get:
\[
k_1= \max\big(\a k_0, \min\big(\b k_0, \frac{ F
  e^{r(T-t_r)}P(t_r)}{S(t_r)}\big)\big)
\]
Note that the reset rule of Eq.~(\ref{eq:3}) is more expensive as the
size of the window of opportunity at a given reset time $t_{r,i}$,
$[\a k_{i-1}S(t_{r,i}),\b k_{i-1}S(t_{r,i})]$, remains constant over
time, while in the case of Eq.~(\ref{eq:4}) this window-size may
shrink over time, thus loosing some value.

In the market, one approximates the price of contracts, with resets
according Eq.~(\ref{eq:4}), as if only the first reset is present
assuming that the stock price at the first reset date will have
dropped so much as to kill all remaining resets.  When this is not the
case the game is repeated but now with the second reset date as the
only reset. Such an approximation will clearly underestimate the price
of the contract, since it neglects the value of the remaining resets.
On the other hand when one would use the reset rule Eq.~(\ref{eq:3})
as a proxy for Eq.~(\ref{eq:4}), the price is overestimated, since the
size of the window of opportunity remains constant over time. A better
proxy might be to take an average of both cases, where both
contributions are weighted with their respective probabilities of
occuring. In a deteriorating market however, the first proxy may not
be such a bad proxy after all.

From section~\ref{sec:definition-problem} it is clear, that the only
way $S$ enters the boundary conditions is through the payoff condition
at maturity and the reset conditions. The boundary conditions for
callable and puttable features remain the same. Now what happens to
the reset conditions? If we can rewrite a reset condition in terms of
the conversion value, then we will be able to put all the effects of
the resets in jump conditions of the following form
\[
v(y,t_-)=v(\max(\a y,\min(\b y,\g),t))
\]
and we do not have to add any extra degree of freedom to solve the
problem. This turns out to be the case, with multiple resets, for the
reset rule of Eq.~(\ref{eq:3}) and, with only one reset
\footnote{For Eq.~(\ref{eq:4}) with multiple resets Monte Carlo
  experiments show that our method serves as a good proxy. We compared
  the European versions of resettable convertibles with the first
  reset condition, a cap $\a=1$, floor $\b=0.8$, and volatility
  $\s=50\%$ gives a difference in price of order $1\%$.  Since the
  second condition is more risky, it will have the lowest price of the
  two. The resets are not really sensitive to callable and puttable
  features, so the difference in prices should also hold for the more
  general case.}.
, for Eq.~(\ref{eq:3}).

\begin{figure}[htbp]
  \begin{center}
    \epsfig{figure=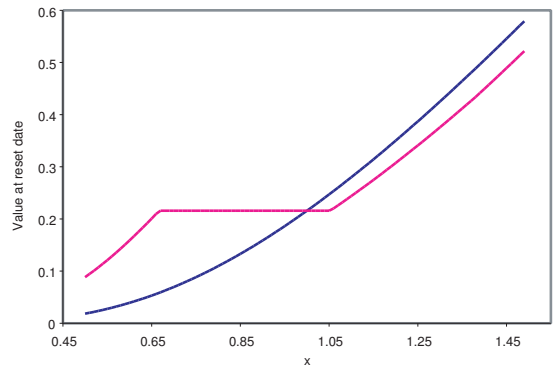,width=7.5cm}
  \end{center}
  \caption{The effect of the jump condition at the reset date $t_1$.}
    \label{fig:1}
\end{figure}

As an example, first consider the simplest case with one reset at time
$t_1$.  There are no callable or puttable features, no coupons and
dividends etc. At maturity the value of the CB is simply
\[
V(x,1,T)=\max\big(k_1 x,F\big)
\]
where we pulled out the numeraire $P$ to simplify the equation, and
introduced $x\equiv S/P$. Note that this is just a function of the
conversion value $y_1\equiv k_1 x$ only. So the value of the contract
at $t_1$ is
\[
f(y_1)= \int \max\big(y_1 \phi(z-\sigma\sqrt{T-t_1}),F\phi(z)\big) dz
\]
of course still a function of $y_1$. Here $\phi$ denotes the standard
normal pdf. The derivation of the above, very useful, formula can be
found in Ref.~\cite{HooglandNeumann99a}. The reset condition of $k_1$
at $t_1$ is given by
\begin{eqnarray*}
  y_1
  &=&\max\big(\a k_0x,\min\big(\b k_0x, F_1\big)\big) \\
  &=&\max\big(\a y_0,\min\big(\b y_0, F_1\big)\big)
\end{eqnarray*}
where $y_0\equiv k_0x$ and $F_1\equiv F\exp(r(T-t_1))$.  At time
$t_1$, $y_1$ is a function of $y_0$ only. The payoff at time $t_1$ in
terms of variable $y_0$ can thus be written as
\[
f(max\big(\a y_0,\min\big(\b y_0, F\big)\big))
\]
This is used as initial value for the PDE in terms of variable $y_0$
until the present time $t$. In Fig.~\ref{fig:1} the adjusted payoff at
time $t_1$ is shown for the case $\a=0.95$ and $\b=1.5$.
In the case where there are coupons, dividends, callable and puttable
features we can use the same trick, but then we have to solve the PDE
for $y_1$ of course numerically from $T$ back to $t_1$.

Now consider the case of multiple resets with the reset condition
specified by
\[
k_ix=\max\big(\a k_{i-1}x,\min\big(\b k_{i-1}x, F_i\big)\big)
\]
where $F_i\equiv F\exp(r(T-t_i))$.  Since the contract only depends on
the products $y_i\equiv k_ix$, we can proceed as above. Starting at
maturity with the payoff
\[
v(y_n,T)=\max(y_n,F)
\]
we solve the PDE backward, in terms of variable $y_n$, in time until
the last reset date $t_{r,n}$.  At that time the value of the contract
is given by $v(y_n,t_{r,n})$.  Now we use the definition of the
conversion ratio reset-rule of $k_n$ at $t_{r,n}$ to rewrite $y_n$ at
time $t_{r,n}$ as a function of the new variable $y_{n-1}\equiv
k_{n-1}x$.
\begin{eqnarray*}
  y_n 
  &=&\max\big(\a k_{n-1},\min\big(\b k_{n-1}, \frac{F_n}{x}\big)\big)x \\
  &=&\max\big(\a y_{n-1},\min\big(\b y_{n-1}, F_n\big)\big)
\end{eqnarray*}
So the value of the contract at time $t_{r,n}$ can now be expressed as
\[
v(max\big(\a y_{n-1},\min\big(\b y_{n-1}, F\big)\big),t_{r,n})
\]
This in turn is a function of $y_{n-1}$ only.  The PDE in terms of
$y_{n-1}$ is identical to the one in terms of $y_n$ and it is solved
backward in time until time $t_{r,n-1}$.  The payoff at time
$t_{r,n-1}$ in terms of variable $y_{n-1}$ can again be rewritten in
terms of $y_{n-2}$. Clearly the procedure can be repeated ad
infinitum.  Also note that the reset rule may be time-dependent.

\section{Resettable CB's are lookbacks in disguise}
\label{sec:nice-relation}

In this section we show that for a particular choice of the reset
there is a close relation between convertibles and lookbacks. This
provides another understanding of why the resets only introduce soft
path-depandancy.  To this end we drop the ceiling ($\b\to\infty$) on
the conversion ratio.  Working out the recursion relation with given
$k_0$, we get for $i=1\ldots n$:
\[
k_i= \max\big( k_0,\max\limits_{j=1\ldots i}e^{r(T-t_{r,j})}\frac{F
  P(t_{r,j})}{S(t_{r,j})} \big)
\]
As we saw earlier the only combination of relevance is the following
expression, defined for $t_{r,i}\le t\le t_{r,i+1}$:
\[
k_i S(t) = \max\big( k_0 S(t), \max\limits_{j=1\ldots
  i}e^{r(T-t_{r,j})}\frac{F P(t_{r,j})}{S(t_{r,j})}S(t) \big)
\]
Using the formulation of Ref.~\cite{HooglandNeumann99b} we can relate
this to the tradables $X_s(t)$ defined through
\[
X_s(t) = \bigg\{
\begin{matrix} 
  P(t) & t<s \\
  \frac{P(s)}{S(s)}{S(t)} & t\ge s
\end{matrix}
\]
If we set the initial conversion ratio at-the-money, $k_0\equiv F
e^{r(T-t_0)}P(t_0)/S(t_0)$, we get
\begin{equation}
  k_i S(t) 
  = \max\limits_{j=0\ldots i} e^{r(T-t_{r,j})}F X_{t_{r,j}}(t) 
\end{equation}
Thus the value of $k_i S(t)$ at times $t\ge t_{r,i}$ is equal to the
value of the weighted maximum of a set of tradables $X_{t_{r,j}}(t)$.
In fact, since we have the freedom to switch numeraires, we can
exchange $S$ and $P$ and with tradable objects $Y_s(t)$ defined
through
\[
Y_s(t) = \bigg\{
\begin{matrix} 
  S(t) & t<s \\
  \frac{S(s)}{P(s)}{P(t)} & t\ge s
\end{matrix}
\]
we can relate the expression with $k_iS$ to
\begin{eqnarray*}
  \hat{k}_i P(t) 
  &=& \max\limits_{j=0\ldots i} e^{r(T-t_{r,j})}F Y_{t_{r,j}}(t) 
\end{eqnarray*}
The $Y_s$ are tradables, that transport the value of $S$ at time $s$
to a later time. So the term $\hat{k}_i$ keeps track of a weighted
maximum of the stock price. Now the governing PDE remains unchanged.
This can be simply understood from the homogeneity of the price as a
function of the tradables. Thus an appropriate change of variables
links a convertible bond with an American lookback.

\section{Results}
\label{sec:results}

We tested the model using various contracts traded on the Japanese
markets. To this end we solved the one-dimensional PDE numerically. In
Ref.~\cite{HooglandNeumann00b} an alternative mixed finite-difference
(FD) scheme, dubbed `tradable scheme' is proposed to solve the PDE in
Eq.~(\ref{eq:1}). It has attractive features compared to the
conventional Crank-Nicholson (CN) scheme and we discuss it shortly
here.  The idea of `tradable schemes' is as follows. We assume that we
are able to solve analytically a given PDE for a given set of simple
boundary conditions. This solution $R$ is then used to construct a
FD-scheme such that the scheme solves $R$ exactly at the grid points.
In contrast to the usual schemes such as CN these schemes behave very
nicely, when there are boundary layer problems, e.g. asian options.
Furthermore it can be formulated in a very compact manner.  More
details can be found in the article mentioned above.

To compute the price of the convertible, using the tradable scheme, we
need a tradable for which we can compute an exact solution. In the
case of deterministic interest rate we can use a power-tradable
\[
R_\a(S,P,\S) = \bigg(\frac{S}{e^{-\half\a\S^2}P}\bigg)^\a
e^{-\half\a\S^2}P
\]
where $\S\equiv\s\sqrt{T-t}$ in the case of constant volatility. We
use $R_2$ to fit the scheme to.

Our results compare favorable with results from commercial packages
for the same parameter settings. Clearly our approach is much faster
and more accurate, than the usual algorithms, due to the reduction of
the statespace.

\begin{figure}[htbp]
  \begin{center}
    \epsfig{figure=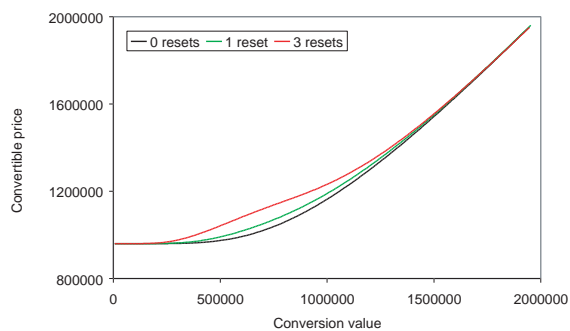,width=8.5cm}
  \end{center}
  \caption{The price of a CB with 0,1, and 3 resets as a function
    of the conversion value.}
  \label{fig:2}
\end{figure}

\begin{figure}[htbp]
  \begin{center}
    \epsfig{figure=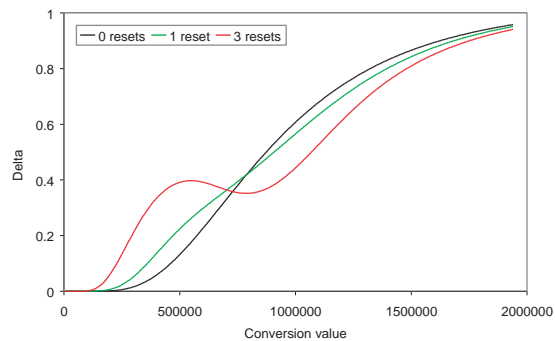,width=7.5cm}
  \end{center}
  \caption{The delta of a CB with 0,1, and 3 resets as a function
    of the conversion value.}
  \label{fig:3}
\end{figure}

\begin{figure}[htbp]
  \begin{center}
    \epsfig{figure=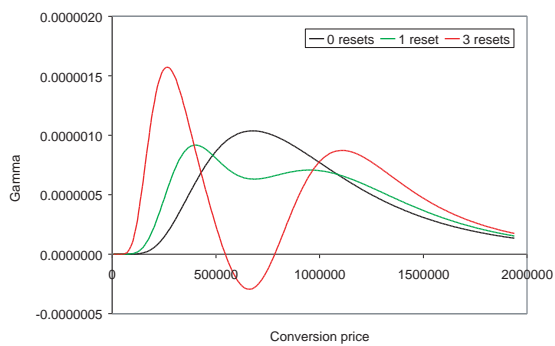,width=7.5cm}
  \end{center}
  \caption{The gamma of a CB with 0,1, and 3 resets as a function
    of the conversion value.}
  \label{fig:5}
\end{figure}

In Figures~\ref{fig:2},~\ref{fig:3}, and ~\ref{fig:5} we have plotted
the price, the delta, and gamma for three CB's differing in the number
of resets. The effect of the resets becomes more pronounced as the
number of resets increases. The resets make the convertible less
sensitive to chan\-ges of the conversion-value in the region where the
contract is allowed to reset.

\begin{figure}[htbp]
  \begin{center}
    \epsfig{figure=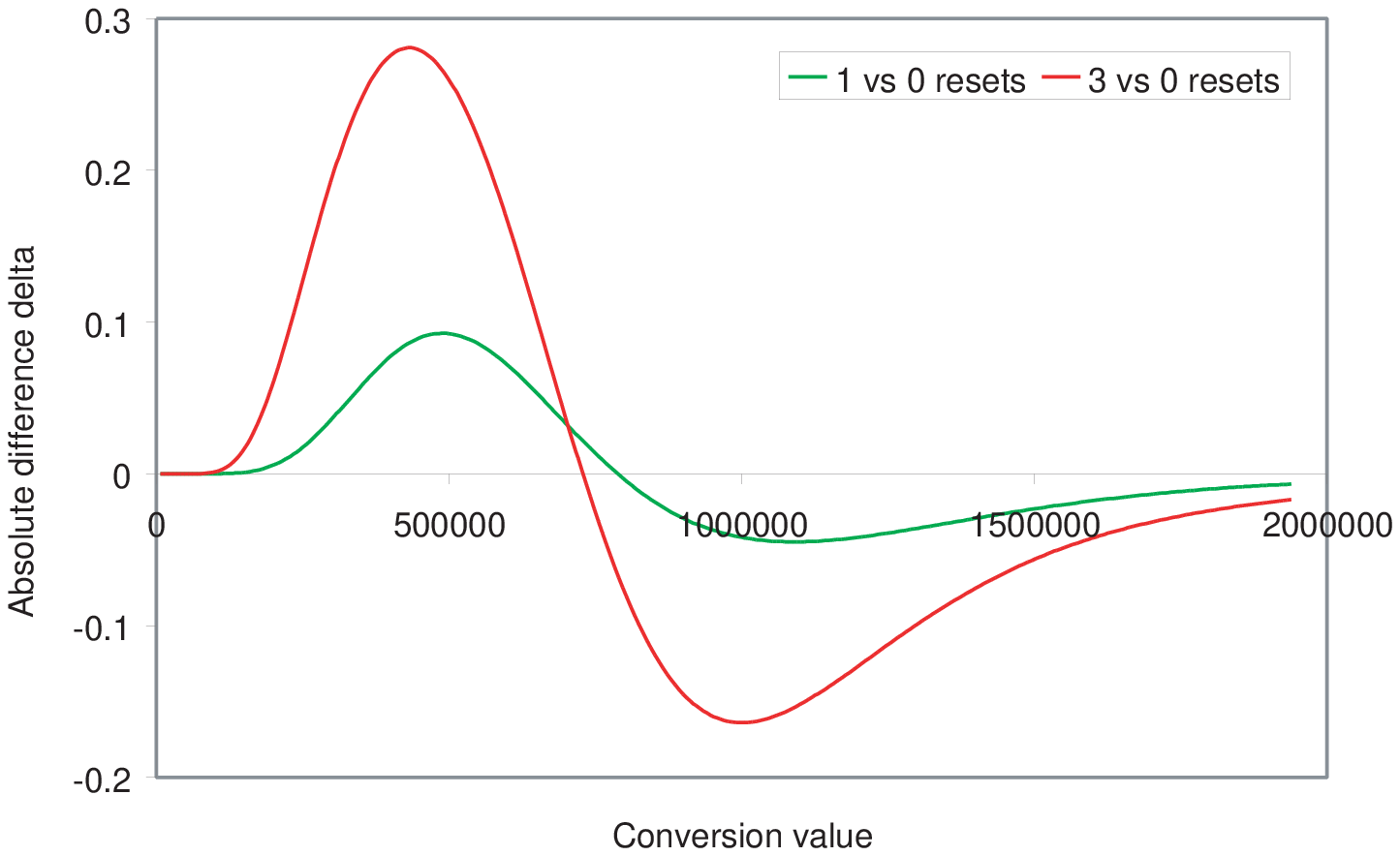,width=7.5cm}
  \end{center}
  \caption{The absolute difference of the delta of a CB with 1 and 3
    resets w.r.t. one without reset as a function of the conversion
    value.}
  \label{fig:4}
\end{figure}

In Figure~\ref{fig:4} the absolute difference of the delta of the two
contracts with resets w.r.t. the one without resets is given. The
impact is quite dramatic. A similar effect can be seen for the gamma's
and hence also for the theta's and vega's. Here one sees in fact that
the gamma may become negative.

When one compares the results with the actual prices in the market,
there is a clear discrepancy. For all contracts we have considered the
model gives a too high price. The implied vol of the contract is much
lower than the historical vol over say the last month or so.  Typical
values are 40\% and 55\% respectively. In a way this signals that one
does not price the embedded option correctly, too cheap. So this might
provide an explanation for the interest of hedge funds in resettable
convertibles.

\begin{figure}[htbp]
  \begin{center}
    \epsfig{figure=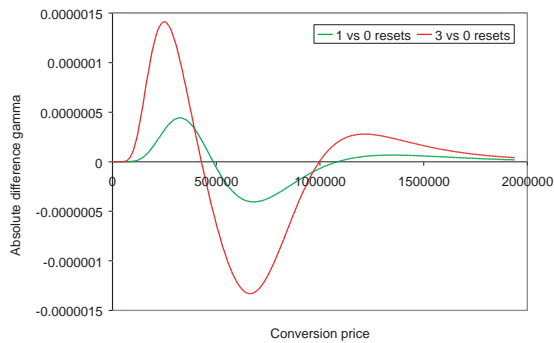,width=7.5cm}
  \end{center}
  \caption{The absolute difference of the gamma of a CB with 1 and 3
    resets w.r.t. one without resets as a function of the conversion
    value.}
  \label{fig:6}
\end{figure}

\section{Conclusions}
\label{sec:conclusions}

We have shown that resets, when using the right coordinates, do not
introduce any extra factor in the PDE used to price the convertible
bond. This significantly reduces the complexity of the problem. In
fact one can show that convertibles are related to lookbacks via a
change of numeraire. A simple algorithm is presented to take into
account the effects of the notice period. All results are given in a
setting with stockprices driven by geometric Brownian motion and
deterministic interest rates, but they carry over too more complex and
realistic situation with stochastic interest rates and credit risk
too. This will be discussed in a follow-up article.

\end{document}